\documentclass[twocolumn,showpacs,preprintnumbers,amsmath,amssymb]{revtex4}
\usepackage{graphicx}
\usepackage{dcolumn}
\usepackage{bm}
\usepackage{feynmf}
\begin{document}
\begin{fmffile}{fmf}
\preprint{APS/123-QED}

\title{Parity inversion, breakdown of shell closure and particle-vibration coupling in Be-isotopes.} 

\author{G.Gori$^{a,b}$, F.Barranco$^{c}$, E.Vigezzi$^{b}$, R.A. Broglia$^{a,b,d}$}

\affiliation{$^a$ Dipartimento di Fisica, Universit\`a degli Studi di Milano,
via Celoria 16, 20133 Milano, Italy.\\
$^b$ INFN, Sezione di Milano, via Celoria 16, 20133 Milano, Italy.\\
$^c$ Departamento de Fisica Aplicada III, Escuela Superior de Ingenieros, Camino de los Descubrimientos s/n,  
41092 Sevilla, Spain.\\
$^d$ The Niels Bohr Institute, University of Copenhagen, Blegdamsvej 17, 2100 Copenhagen \O, Denmark.
}

\date{\today}

\begin{abstract}
The coupling of single-particle motion and of vibrations in $^{11}_4$Be produces 
dressed neutrons which spend only a fraction of the time in pure single-particle 
states, and which weighting differently from the bare neutrons, lead to parity inversion.
The interaction of the two least bound neutrons in the ground state of $^{12}_4$Be mediated by
the $v_{14}$ Argonne nucleon-nucleon potential and by the exchange of the phonon cloud, give
rise to a strongly correlated state, where the neutrons spend more than half of the time in 
(s$^2$+d$^2$)-configurations, resulting in a breaking of the $N=8$ shell closure. 
\end{abstract}


\maketitle
 
The properties of finite many-body systems are strongly influenced by spatial quantization \cite{Kub.62}
leading to marked shell structures \cite{BMI:69}.
This type of quantal size effects are, on the other hand, renormalized in an important way
by zero point fluctuations \cite{BMII:75,BerBro:91}. The smaller the system is, the largest the surface/volume ratio is 
and the strongest these effects may become. In particular, the interweaving of single-particle motion
and of collective
vibrations of the surface of atomic nuclei can lead to state dependent effective masses which, for light systems
can even invert the sequence of particular levels, thus altering the magic numbers associated
with closed shell.
While the standard parametrization of the single-particle potential \cite{BMI:69} indicates the states 0p$_{1/2}$ and 
1s$_{1/2}$ to be the last bound and the first unbound orbitals of the system with $Z=4$ and $N=7$ nucleons, 
$^{11}_4$Be$_7$ displays parity inversion with respect to this sequence of single particle levels. In fact, the two 
bound states of this system have quantum numbers (binding energy) 1/2$^+$ (-0.5 MeV) and 1/2$^-$
(-0.18 MeV) respectively \cite{Mil.83}. The associated spectroscopic factors
measured in the knock-out reaction $^{12}$Be($^9$Be,$^{9}$Be+n+$\gamma$)$^{11}$Be 
are 0.42$\pm$0.06 and 0.37$\pm$0.06 respectively 
\cite{Nav.00}, indicating that $N=8$ is not a good magic number in the case of the neutron rich nucleus $^{12}$Be,
as it was not in the case of $^{11}$Li either. In connection with this discussion, cf. also \cite{Tal60}.

Making use of the fact that the effective mass ($\omega$-mass) $m_{\omega}=m(1+\lambda)$ \cite{Mah.85}
arising from the coupling to the collective vibrations
of the system as measured by the mass enhancement factor $\lambda$, is closely connected with
the associated spectroscopic factor $Z_{\omega}=m/m_{\omega}=(1+\lambda)^{-1}$ of
levels close to the Fermi energy, and of the experimental finding quoted above ($Z_{\omega}\approx 0.3-0.5$) one estimates 
$\lambda\approx 1.2$. Making use of this value, the two-particle separation energy 
associated to the least bound neutron of $^{12}$Be is expected to be $W\approx -\lambda \omega_C$
(strong coupling limit of the Cooper pair solution \cite{Sch.64}), where $\omega_C$ is a typical energy of vibrational 
modes, ($\approx 3.37$ MeV, quadrupole vibration in $^{10}$Be \cite{Iwa.00}). 
Consequently $W\approx -4$MeV, a value which agrees
pretty well with the experimental finding ($S_{2n}=-3.7$ MeV \cite{Aud.95}). From these simple estimates, 
one expects the renormalization effects of the nucleon-nucleon interaction (induced interaction) of
the two least bound neutrons moving around the core $^{10}$Be due to the 
particle-vibration phenomenon, to be conspicuous.  

In what follows we shall study the nuclei $^{11}_4$Be$_7$ and $^{12}_4$Be$_8$, within the framework of
nuclear field theory ($NFT$) \cite{bes1,bes2,Bro.76,bes3,Bor.77,Rei.75}, allowing the nucleons 
to interact through a nucleon-nucleon realistic potential (Argonne $v_{14}$ \cite{Wir.84}) taking 
also into account the coupling between single-particle motion and collective vibrations of the system. 
Special emphasis will be set, in the present case, in the $NFT$ calculation of the spectroscopic 
factors of $^{12}$Be which, together with the the ground state occupation probabilities 
of the two-particle configurations $s^2$, $p^2$ and $d^2$, 
provide the most sensitive predictions for a detailed comparison with the experimental findings.

We start by considering the system $^{11}_4$Be$_7$ described as one neutron moving around the core 
$^{10}_4$Be$_6$, in keeping with the fact that the value of the neutron separation energy in $^{10}$Be
is 6.813 MeV as compared with the value of 0.504 MeV in $^{11}$Be (\cite{Aud.95}). 
The single-particle levels are determined by solving the Schr\"{o}dinger equation
\begin{equation}
\left(-\frac{\hbar^2}{2m_k}\nabla^2_r+U'(r) \right) \phi_j(r)=\epsilon_j \phi_j(r),
\label{eqn:schr}
\end{equation}
in a spherical box of radius equal to 30 fm so as to discretized the continuum states.  
The quantity $m_k$ is the $k$-mass while $U'(r)=(m/m_k) U(r)$, $U(r)$ being a Saxon-Woods potential 
with a standard parameterization for the depth \cite{BMI:69} 
\begin{equation}
V=-50.5+33 \frac{N-Z}{A}\rm{ MeV.}
\end{equation}

In keeping with the fact that the $k$-mass is directly connected with non-locality effects 
(mainly exchange effects associated with the Fock potential),
it is expected to strongly depend on the density of the system \cite{Mah.85}. 
In the case of nuclei along the stability valley $m_k\approx 0.7 m$, 
while in the case of halo nuclei like $^{11}$Be, one expects $0.8m \leq m_k \leq m$.
Calculations using both of the limiting values of $m_k$ were carried out, with rather similar results, as explained below.

Making use of the associated particle-hole basis and of a separable multipole-multipole interaction, 
the $L^{\pi}=2^+$ and $3^-$ vibrations were calculated in the random phase approximation ($RPA$).
A self-consistent coupling constant $k_L$ \cite{BMII:75}, 
slightly adjusted to reproduce the energy of quadrupole vibrations, was used. 
This is possible for both $m_k/m=1$ and $m_k/m=0.8$  in keeping with the fact that
$k_L$ is proportional to the density weighted matrix element of $\partial U'/\partial r$,
a value which scales with ($m/m_k$) (cf. Eq.(\ref{eqn:schr})).
The range of the associated deformation parameters $\beta_L$ is consistent with observation \cite{Iwa.00,Ram.87}.

\begin{figure}[t]
\parbox{80mm}{
\begin{tabular}{|c| c|}
\cline{2-2}
\multicolumn{1}{c|}{}  & \\
\multicolumn{1}{c|}{}  &  $\parbox{8mm}{
\begin{fmfgraph*}(10,10)
  \fmftop{ou1}
  \fmfbottom{in1}
    \fmf{fermion}{in1,ou1}
\end{fmfgraph*}}(j\pi)$ \\
\multicolumn{1}{c|}{}  & \\
\hline
& \\
$\parbox{8mm}{
\begin{fmfgraph*}(10,10)
  \fmftop{ou1}
  \fmfbottom{in1}
    \fmf{fermion}{in1,ou1}
\end{fmfgraph*}}(j\pi)$ &
$\parbox{8mm}{
\begin{fmfgraph*}(10,10)
  \fmftop{ou1}
  \fmfbottom{in1}
    \fmf{fermion}{in1,ou1}
\end{fmfgraph*}}\ \ $+
\parbox{20mm}{
\begin{fmfgraph*}(20,20)
  \fmftop{in2}
  \fmfbottom{in1}
    \fmf{fermion}{in1,v1}
    \fmf{fermion}{v1,v2}
    \fmf{fermion}{v2,in2}
    \fmf{boson,tension=0.5,right=0.8}{v1,v2}
    \fmfforce{.5w,.25h}{v1}
    \fmfforce{.5w,.75h}{v2}
    \fmfforce{.5w,0h}{i1}
    \fmfforce{.5w,1h}{i2}
\end{fmfgraph*}}+
$\parbox{15mm}{\begin{fmfgraph*}(15,20)
  \fmftop{i2}
  \fmfbottom{i1}
    \fmf{fermion}{i1,v1}
    \fmf{fermion}{v1,v2}
    \fmf{fermion}{v2,i2}
    \fmf{boson,tension=0.5,left=0.5}{v1,v2}
    \fmfforce{0w,.6h}{v1}
    \fmfforce{1w,.3h}{v2}
    \fmfforce{0w,0h}{i1}
    \fmfforce{1w,1h}{i2}
\end{fmfgraph*}} \ \ \ $
\\
& ~~~~~~~ (a)~~~~~~~~~~~~~~~(b)~~~~~~~~~~~~~~~~~(c) \\\hline
\end{tabular}
}
\caption{Schematic representation of the effective matrix
used in the Bloch-Horowitz perturbation theory to calculate  the eigenvalues of $^{11}$Be.
An arrowed line pointing upwards (downwards) indicate a particle (hole), while a wavy 
line indicate a collective vibrational state} \label{fig:1pmat}
\end{figure}
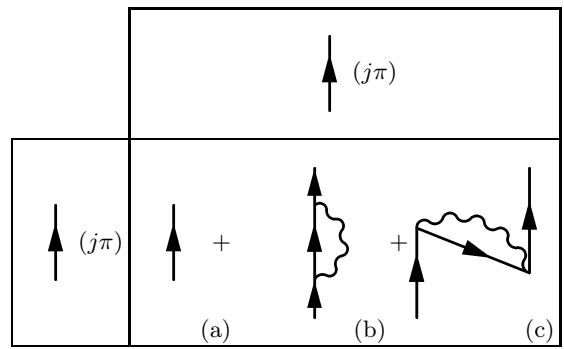

The interweaving of single-particle motion and of collective degrees of freedom arising
from the particle-vibration Hamiltonian and from the bare nucleon-nucleon interaction $v_{14}$
has been treated within the framework
of the $NFT$ making use of Bloch-Horowitz perturbation theory \cite{Bor.77,Dus.71}.
The eigenvalues of the dressed single-particle states were obtained
by diagonalizing (energy dependent) matrices of the order 10$^2\times \ 10^2$ whose elements
connect a basis of unperturbed states containing both bound and continuum 
solutions of Eq. (\ref{eqn:schr}) with energies up to 350 MeV, with states containing a particle
and a vibration (Fig. (\ref{fig:1pmat}b)) as well as two particles 
and a hole plus a collective mode (Fig. (\ref{fig:1pmat}c)). The calculation were carried out 
for states with quantum number s$_{1/2}$, p$_{1/2}$ and d$_{5/2}$. 
Similar results were obtained making use of the unperturbed 
single-particle basis calculated solving Eq. (\ref{eqn:schr}) with $m_k/m$=1 and 
$m_k/m$=0.8, as the larger (absolute) values of the energies $\epsilon_j$ are compensated by the stronger
particel-vibration coupling vertices  proportional to $\beta_L$ and to $\partial U'/\partial r$.
In what follows we shall refer to the results obtained with $m_k/m$=1, results which are displayed
in Table \ref{tab1}, in comparison with the experimental findings.
Theory provides an overall account of the experimental findings, 
also concerning the spectroscopic factors associated with the reaction $^{10}$Be(d,p)$^{11}$Be \cite{Au.70}.
The way these quantities were calculated are discussed below in connection with a similar calculation
carried out in connection with the reaction $^{12}$Be($^9$Be,$^9$Be+$n+\gamma$)$^{11}$Be \cite{Nav.00}.

\begin{table}[h]
\begin{tabular}{|c|c|c|c|c|}  \cline{3-5}
\multicolumn{1}{c}{}  &   &         &   \multicolumn{2}{c|}{}       \\
\multicolumn{1}{c}{}  &   &         &   \multicolumn{2}{c|}{Theory}       \\
\multicolumn{1}{c}{}  &   &         &   \multicolumn{2}{c|}{}       \\  \cline{4-5}
\multicolumn{1}{c}{}  &   &  &  &       \\
\multicolumn{1}{c}{}  &   &  \multicolumn{1}{c|}{ Exper.}& \multicolumn{1}{c|}{particle-vibration} & mean field      \\
\multicolumn{1}{c}{}  &   &  &  &       \\ \hline
                   &    &                  &            &              \\
                   &  E$_{s_{1/2}}$ &  -0.504 MeV     &   -0.48 MeV  & $\sim$ 0.14 MeV \\
                   &    &                  &            &              \\   \cline{2-5}
                   &    &                  &            &              \\
       &  E$_{p_{1/2}}$ &  -0.18 MeV     &   -0.27 MeV  & -3.12 MeV \\
                   &    &                  &   &     \\ \cline{2-5}
                   &    &                  &            &              \\
$^{11}_4$Be$_7$       &  E$_{d_{5/2}}$ &       &   $\sim$ 0 MeV  & $\sim$ 0.7 MeV \\
    &    &                  &   &     \\ \cline{2-5}
                   &    &                  &            &           \\
       &  S$\left[1/2^+\right]$ &   0.73 $\pm$0.06    &   0.76   &0  \\
                   &    &                  &            &     \\ \cline{2-5}
                   &    &                  &            &           \\
       &  S$\left[1/2^-\right]$ &  0.63$\pm$0.15      &   0.79  &  1 \\
                   &    &                  &     &     \\ \hline
             &         &                          &             &         \\
             &S$_{2n}$ &  -3.673  MeV     &   -3.58 MeV  & -6.24 MeV \\
             &         &                          &             &         \\ \cline{2-5}
             &         &                          &             &          \\
$^{12}_4$Be$_8$    & s$^2$,p$^2$,d$^2$   &   & 23\%, 29\%, 48\%  &0\%,100\%,0\%\\
             &         &                          &             &           \\   \cline{2-5}
                   &               &              &            &           \\
&  S$\left[1/2^+\right]$  &  0.42$\pm$0.10      &   0.27  &  0\\
             &         &                          &               &         \\ \cline{2-5}
                   &               &              &            &           \\
 & S$\left[1/2^-\right]$  &  0.37$\pm$0.10
&   0.32  & 1\\
             &         &                          &               &         \\ \hline
\end{tabular} 
\caption{Comparison of experimental binding energy and
spectroscopic factors with those resulting from our calculation
and from an independent particle model.The spectroscopic factors are those for the transfer of one particle
on s$_{1/2}$ and p$_{1/2}$ states and they are measured for $^{11}$Be and $^{12}$Be in [21] and [6]
respectively. For $^{12}$Be, 
we also show the
components of the resulting ground state wave function.}
\label{tab1} 
\end{table}

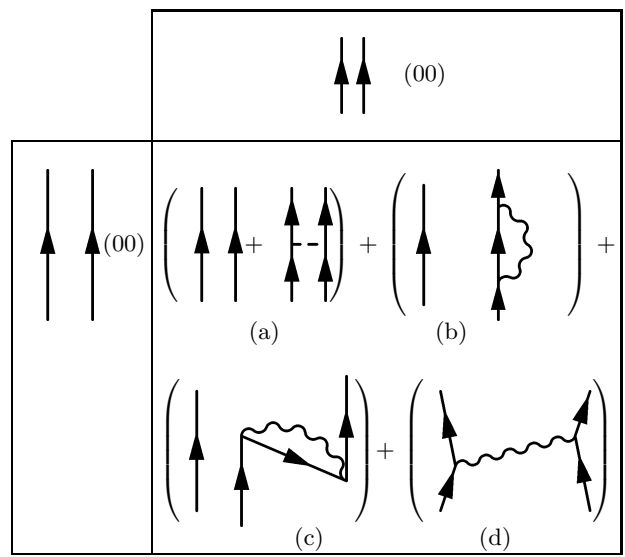
\begin{figure}[t]
\parbox{80mm}{
\begin{tabular}{|c| c|}
\cline{2-2}
\multicolumn{1}{c|}{}  & \\
\multicolumn{1}{c|}{}  &  $\parbox{8mm}{
\begin{fmfgraph*}(10,10)
  \fmftop{ou1,ou2}
  \fmfbottom{in1,in2}
    \fmf{fermion}{in1,ou1}
    \fmf{fermion}{in2,ou2}
\fmfforce{0.2w,0h}{in1}
\fmfforce{0.2w,1h}{ou1}
\fmfforce{0.5w,0h}{in2}
\fmfforce{0.5w,1h}{ou2}
\end{fmfgraph*}} \ \ (00)$ \\
\multicolumn{1}{c|}{}  & \\
\hline
& \\
$ \parbox{8mm}{
\begin{fmfgraph*}(20,20)
  \fmftop{ou1,ou2}
  \fmfbottom{in1,in2}
    \fmf{fermion}{in1,ou1}
    \fmf{fermion}{in2,ou2}
\fmfforce{0.2w,0h}{in1}
\fmfforce{0.2w,1h}{ou1}
\fmfforce{0.5w,0h}{in2}
\fmfforce{0.5w,1h}{ou2}
\end{fmfgraph*}} \ \ \  (00)$ &
$\left( \parbox{8mm}{
\begin{fmfgraph*}(15,15)
  \fmftop{ou1,ou2}
  \fmfbottom{in1,in2}
    \fmf{fermion}{in1,ou1}
    \fmf{fermion}{in2,ou2}
\fmfforce{0.2w,0h}{in1}
\fmfforce{0.2w,1h}{ou1}
\fmfforce{0.5w,0h}{in2}
\fmfforce{0.5w,1h}{ou2}
\end{fmfgraph*}} 
+\parbox{8mm}{
\begin{fmfgraph*}(15,15)
  \fmftop{ou1,ou2}
  \fmfbottom{in1,in2}
    \fmf{fermion}{in1,v1}
\fmf{fermion}{v1,ou1}
    \fmf{fermion}{in2,v2}
\fmf{fermion}{v2,ou2}
\fmf{dashes}{v1,v2}

\fmfforce{0.2w,0h}{in1}
\fmfforce{0.2w,1h}{ou1}
\fmfforce{0.5w,0h}{in2}
\fmfforce{0.5w,1h}{ou2}
\fmfforce{0.2w,0.5h}{v1}
\fmfforce{0.5w,0.5h}{v2}
\end{fmfgraph*}} 
\right)$ + 
$\left( \ \  \parbox{18mm}{
\begin{fmfgraph*}(20,20)
  \fmftop{ou2,in2}
  \fmfbottom{ou1,in1}
    \fmf{fermion}{ou1,ou2}
    \fmf{fermion}{in1,v1}
    \fmf{fermion}{v1,v2}
    \fmf{fermion}{v2,in2}
    \fmf{boson,tension=0.5,right=0.8}{v1,v2}
    \fmfforce{.5w,.25h}{v1}
    \fmfforce{.5w,.75h}{v2}
    \fmfforce{.5w,0h}{in1}
    \fmfforce{.5w,1h}{in2}
\end{fmfgraph*}} \ \right) $ +  \\
& (a)~~~~~~~~~~~~~~~~~~~(b)~~~~~~~~\\
& \\
& $\left( \ \parbox{22mm}{\begin{fmfgraph*}(20,20)
  \fmftop{p1,i2}
  \fmfbottom{p2,i1}
    \fmf{fermion}{p2,p1}
    \fmf{fermion}{i1,v1}
    \fmf{fermion}{v1,v2}
    \fmf{fermion}{v2,i2}
    \fmf{boson,tension=0.0,left=0.5}{v1,v2}
    \fmfforce{0.3w,.6h}{v1}
    \fmfforce{1w,.3h}{v2}
    \fmfforce{0.3w,0h}{i1}
    \fmfforce{1w,1h}{i2}
\end{fmfgraph*}} \right) $+
$\left( \ \  \parbox{18mm}{\begin{fmfgraph*}(20,20)
  \fmftop{ou1,ou2}
  \fmfbottom{in1,in2}
    \fmf{fermion}{in1,v1}
    \fmf{fermion}{in2,v2}
    \fmf{fermion}{v2,ou2}
    \fmf{fermion}{v1,ou1}
    \fmf{boson}{v1,v2}
    \fmfforce{.1w,.4h}{v1}
    \fmfforce{.9w,.6h}{v2}
\end{fmfgraph*}}  \ \ \right)$ \\
& ~~~~~~(c)~~~~~~~~~~~~~~~~~~~(d)~~~\\
\hline
\end{tabular}
}
\caption{Schematic representation of the effective matrix
used in the Bloch-Horowitz perturbation theory to calculate  the eigenvalues of $^{12}$Be.
The dashed horizontal line represent the bare nucleon-nucleon potential.
Pairs of nucleons are coupled to angular momentum $L=0$.
} \label{fig:effmat12}
\end{figure}

The self-energy processes used to describe the  dressed single-particle states of $^{11}$Be which eventually
accounted for the parity inversion experimentally observed, 
have been included in the description of the ground state
of $^{12}$Be as it can be seen from Fig. \ref{fig:effmat12}
which shows the effective matrix to be diagonalized in order
to describe the ground-state properties of the correlated three body system $^{12}$Be.
The Hilbert space used to describe $^{12}$Be is
made out of two particle states (cf. Fig. \ref{fig:effmat12}(a)), two particles and one phonon 
(Fig. \ref{fig:effmat12}(b)+Fig. \ref{fig:effmat12}(d)), and three
particles-one hole and one phonon states (Fig. \ref{fig:effmat12}(c)).
All these configurations are coupled to zero angular momentum
and display energies up to 500 MeV.
The effects of $v_{14}$ and of the particle-vibration coupling in $^{12}$Be are calculated by diagonalizing
the effective, energy dependent ($\approx$ 10$^3\times 10^3$) matrix built on a basis of two particles coupled to zero angular momentum.
The lowest eigenvalue -3.58 MeV is to be compared with
the experimental two
particle separation energy of -3.67 MeV. 
The main contribution to the nucleon-nucleon interaction arises
from the induced interaction (Fig. \ref{fig:effmat12}(d)), that 
associated with the bare nucleon-nucleon potential (cf. Fig. \ref{fig:effmat12}(a))
being very small.
This result is associated with the fact
that, in $^{12}$Be, the single-particle states allowed to the valence
neutrons to correlate are essentially the s$_{1/2}$, p$_{1/2}$ and d$_{5/2}$
orbits. Consequently the two neutrons are not able, in this
low-angular momentum phase space, to profit from the
pairing contribution of the bare interaction which is connected with
the high components of the associated multipole expansion.

The squared amplitude of the $^{12}$Be ground state wavefunction are shown in table 
\ref{tab1}.
The large d$^2_{5/2}$(0)-amplitude
predicted for the $^{12}$Be ground state (cf. also \cite{Nav.00}) as compared to that
calculated in the case of $^{11}$Li \cite{Bar.01} can be understood in terms of
the fact that the d$_{5/2}$ orbital is, in $^{10}$Li much less confined than
in $^{11}$Be, thus displaying much smaller overlaps with the 1s$_{1/2}$
and 0p$_{1/2}$ orbitals. Furthermore to the fact that in $^{11}$Li 
the dipole mode is much softer than in $^{12}$Be \cite{Iwa.00b}.
Using this wave function and those
obtained for the ground state and the first excited state of $^{11}$Be,
we have calculated the spectroscopic factors associated to
the knock-out of a single nucleon. In particular, a nucleon with 
quantum number $1/2^-$, 
$$
S[1/2^-]=|\langle ^{11}\mbox{Be}|\mbox{a}_{1p_{1/2}}|^{12}\mbox{Be} \rangle|^2=|T_{1/2^-}|^2.
$$
In order to do this, used is made of 
the eigenvectors of the effective matrix associated with the lowest excited state of $^{11}$Be and 
with the ground state of $^{12}$Be. We will indicate the former with $\tilde{\xi}_p$ (the index $p$
indicating the p$_{1/2}$ single-particle states of the basis) and the latter with $\tilde{\xi}_{ii}$
(the index $ii$ indicating that the corresponding basis state contains a pair of particles moving in the level $i$). 
The components
$\tilde{\xi}_{p}$ and $\tilde{\xi}_{ii}$ are related to the physical amplitudes $\xi_p$ and $\xi_{ii}$
through the relations (cf. e.g. \cite{Bor.77})
$$
\xi_p=\tilde{\xi}_p/\sqrt{N(^{11}{\rm Be})}, ~~~~~~~~~~~~~ \xi_{ii}=\tilde{\xi}_{ii}/\sqrt{N(^{12}{\rm Be})},
$$
where $N(^{11}{\rm Be})$ and $N(^{12}{\rm Be})$ are the normalization factors for $^{11}$Be and $^{12}$Be respectively.
In terms of Feynman diagrams $N(^{12}{\rm Be})$ is given by
\begin{widetext}
$$
N(^{12}{\rm Be})=\sum_{
ii'
}
\tilde{\xi}^2_{ii'}+
\sum_{
\begin{array}{c}
ii'\\
jj'
\end{array}
}
\tilde{\xi}_{ii'} \tilde{\xi}_{jj'}\times \left\{
\sum_{\lambda}
\parbox{20mm}{
\begin{fmfgraph*}(25,25)
  \fmftop{ou1,ou2}
  \fmfbottom{in1,in2}
    \fmf{fermion}{in2,v2}
    \fmf{fermion}{v2,ou2}
    \fmf{fermion,label=$i'$}{in1,v1}
    \fmf{fermion,label=$j'$,l.s=right}{v1,ou1}
    \fmf{boson,label=$\lambda$,l.s=right}{v1,v2}
\fmfforce{0.25w,.4h}{v1}
\fmfforce{0.75w,.6h}{v2}
\fmfforce{0w,.9h}{ou1}
    \fmfv{label=$j$,l.a=-30}{ou2}
    \fmfv{label=$i$,l.a=30}{in2}
    \fmfforce{0.9w,0.5h}{p1}
    \fmfforce{0.1w,0.5h}{p2}
    \end{fmfgraph*}}\right. \ \ \ \ \ \ \ 
+\sum_{m,\lambda}\ \ \ 
\parbox{20mm}{
\begin{fmfgraph*}(25,25)
  \fmftop{ou1,ou2}
  \fmfbottom{in1,in2}
    \fmf{fermion}{in2,ou2}
    \fmf{fermion,label=$i'$}{in1,v1}
    \fmf{fermion,label=$m$,l.s=left}{v1,v2}
    \fmf{fermion,label=$j'$}{v2,ou1}
    \fmf{boson,tension=0.5,right=0.5,label=$\lambda$,l.s=right}{v1,v2}
\fmfv{label=$j$,l.a=-30}{ou2}
\fmfv{label=$i$,l.a=30}{in2}
\fmfforce{0.2w,0h}{in1}
\fmfforce{0.7w,0h}{in2}
\fmfforce{0.2w,.2h}{v1}
\fmfforce{0.2w,.7h}{v2}
\fmfforce{0.7w,1h}{ou2}
\fmfforce{0.2w,1h}{ou1}
    \fmfforce{0.9w,0.6h}{p1}
    \fmfforce{0.1w,0.6h}{p2}
    \end{fmfgraph*}}
+\sum_{\lambda}
\left.  \ \
\parbox{20mm}{
\begin{fmfgraph*}(25,25)
  \fmftop{ou2,in1}
  \fmfbottom{in2,ou1}
    \fmfforce{0.45w,0h}{in2}
    \fmfforce{0.2w,0.8h}{in1}
    \fmfforce{0.45w,0.8h}{ou2}
    \fmfforce{0.2w,0h}{ou1}
    \fmfforce{0.45w,0.5h}{v3}
    \fmfforce{0.65w,0.8h}{b2}
    \fmfforce{0.65w,0.2h}{b1}
    \fmf{fermion}{in2,b2}
    \fmf{fermion,tension=0.8,left=1,label=$p_{3/2}$,l.s=left}{b2,b1}
    \fmf{vanilla,l.s=right}{b1,v3}
    \fmf{fermion}{v3,ou2}
    \fmf{fermion}{ou1,in1}
    \fmf{boson,label=$\lambda$,l.s=right}{b1,b2}
    \fmfv{label=$j'$,l.a=-135}{in1}
    \fmfv{label=$i'$,l.a=135}{ou1}
    \fmfv{label=$i$,l.a=30}{in2}
    \fmfv{label=$j$,l.a=-30}{ou2}
    \fmfforce{1.1w,0.6h}{p1}
    \fmfforce{0.1w,0.6h}{p2}
\end{fmfgraph*}}\ \ \ \ \ \  \ \ \ \ \ \ \ 
\right\}
$$
\end{widetext}
Requiring that $\sum_{ii} \tilde{\xi}^2_{ii}=1$, one obtains 
$N(^{12}{\rm Be})$=0.47. The fact that $N$ is smaller than one, 
and consequently that the amplitudes $\xi_{ii}$ are larger than unity,
is associated with the non-hermitian character of the effective Hamiltonian used in
the calculations. This property is a consequence of 
the overcompleteness of the basis used in $NFT$ (containing both single-particle and collective degrees of freedom)
as well as of the presence of Pauli principle violating components.
The calculation of the normalization factor $N(^{11}{\rm Be})$ carried out using a Feynman
diagram expansion
similar to that used for $^{12}$Be leads to a value of 0.48. By using the physical amplitudes $\xi_p$ and 
$\xi_{ii}$ the value of $T_{1/2^-}$ is obtained through the evalutation  of
\begin{widetext}
$$
T_{1/2^-}=
\left\{\sum_{
\begin{array}{c}
p\\
pp'\\
\end{array}
}
\xi_{p} \xi_{pp'}\times
\parbox{20mm}{
\begin{fmfgraph*}(25,25)
  \fmfbottom{ou1,ou2}
  \fmftop{in2}
    \fmfforce{0.1w,1h}{in2}
    \fmfforce{0.1w,0h}{ou2}
    \fmfforce{0.35w,0h}{ou1}
    \fmfforce{0.6w,0.5h}{v1}
    \fmfforce{0.35w,0.5h}{v2}
    \fmf{fermion}{ou2,in2}
    \fmf{fermion}{ou1,v2}
    \fmf{dashes,label=$a_{1p_{1/2}}$,l.s=left}{v2,v1}
    \fmfv{label=$p$,l.a=0}{in2}
    \fmfv{label=$pp'$,l.a=0}{ou1}
\end{fmfgraph*}} 
+\sum_{
\begin{array}{c}
p,\lambda\\
dd'\\
\end{array}
}\xi_{p} \xi_{dd'}\times
\parbox{27mm}{
\begin{fmfgraph*}(25,25)
  \fmftop{ou1,ou2}
  \fmfbottom{in1,in2}
    \fmf{fermion}{in2,v2}
    \fmf{fermion}{v2,ou2}
    \fmf{fermion}{in1,v1}
    \fmf{fermion}{v1,b1}
    \fmf{boson,label=$\lambda 3^-$,l.s=right}{v1,v2}
\fmf{dashes,label=$a_{1p_{1/2}}$,l.s=left}{ou1,b1}
\fmfforce{0.25w,.35h}{v1}
\fmfforce{0.75w,.65h}{v2}
\fmfforce{0.15w,.6h}{b1}
\fmfforce{0w,.6h}{ou1}
    \fmfv{label=$dd'$,label.angle=180}{in2}
    \fmfv{label=$p$,label.angle=135}{ou2}
    \end{fmfgraph*}} \right.
+\sum_{
\begin{array}{c}
p'',\lambda\\
m,pp'\\
\end{array}
}\xi_{p''} \xi_{pp'}\times \ \ \ \ \  \ \ \
\parbox{27mm}{
\begin{fmfgraph*}(25,25)
  \fmftop{ou1,ou2}
  \fmfbottom{in1,in2}
    \fmf{fermion}{in2,b1}
    \fmf{fermion}{in1,v1}
    \fmf{fermion}{v1,v2}
    \fmf{fermion}{v2,ou1}
    \fmf{boson,tension=0.5,right=0.5,label=$\lambda 3^-$,l.s=right}{v1,v2}
\fmf{dashes,label=$a_{1p_{1/2}}$}{ou2,b1}
\fmfv{label=$p''$,l.a=0}{ou1}
\fmfv{label=$pp'$,l.a=0}{in2}
\fmfv{label=$m d_{5/2}$,l.a=180}{v1}
\fmfforce{0w,0h}{in1}
\fmfforce{0.7w,0h}{in2}
\fmfforce{0w,.2h}{v1}
\fmfforce{0w,.7h}{v2}
\fmfforce{.7w,.7h}{b1}
\fmfforce{1w,.7h}{ou2}
    \end{fmfgraph*}}
$$
$$
(1.01)~~~~~~~~~~~~~~~~~~~~~~~~~~~~(0.01)~~~~~~~~~~~~~~~~~~~~~~~~~~~(0.01)
$$
$$
+\sum_{
\begin{array}{c}
p'',\lambda\\
pp'\\
\end{array}
}\xi_{p''} \xi_{pp'}\times
\left. \left[ \ \
\parbox{30mm}{
\begin{fmfgraph*}(25,25)
  \fmftop{ou2}
  \fmfbottom{in2,ou1}
    \fmfforce{0.45w,0h}{in2}
    \fmfforce{0.45w,1h}{ou2}
    \fmfforce{0.2w,0h}{ou1}
    \fmfforce{0w,0.5h}{v1}
    \fmfforce{0.2w,0.5h}{v2}
    \fmfforce{0.45w,0.5h}{v3}
    \fmfforce{0.65w,0.8h}{b2}
    \fmfforce{0.65w,0.2h}{b1}
    \fmf{fermion}{in2,b2}
    \fmf{fermion,tension=0.8,left=0.8,label=$1p_{3/2}$,l.s=left}{b2,b1}
    \fmf{fermion,l.s=right}{b1,v3}
    \fmf{fermion}{v3,ou2}
    \fmf{fermion}{ou1,v2}
    \fmf{dashes,label=$a_{1p_{1/2}}$}{v2,v1}
    \fmf{boson}{b1,b2}
    \fmfv{label=$p''$,l.a=180}{ou2}
    \fmfv{label=$pp'$,l.a=180}{ou1}
    \fmfv{label=$\lambda 2^+$,l.a=-90}{b1}
\end{fmfgraph*}} \ \ \ +\ \ \parbox{30mm}{
\begin{fmfgraph*}(25,25)
  \fmftop{ou1,ou2}
  \fmfbottom{in2,in1}
    \fmfforce{1.w,0h}{in2}
    \fmfforce{0.2w,0h}{in1}
    \fmfforce{0.45w,.5h}{v2}
    \fmfforce{0.2w,1h}{ou1}
    \fmfforce{1w,0.7h}{v1}
    \fmfforce{0.75w,0.7h}{v2}
    \fmfforce{0.75w,0.5h}{v3}
    \fmfforce{0.55w,0.8h}{b2}
    \fmfforce{0.55w,0.2h}{b1}
    \fmf{fermion,l.s=right}{in2,b2}
    \fmf{fermion,tension=0.8,right=0.8}{b2,b1}
    \fmf{vanilla,l.s=right}{b1,v3}
    \fmf{fermion,l.s=right}{v3,v2}
    \fmf{fermion}{in1,ou1}
    \fmf{dashes,label=$a_{p_{1/2}}$}{v1,v2}
    \fmf{boson}{b1,b2}
    \fmfv{label=$pp'$}{in1}
    \fmfv{label=$p''$,l.a=180}{ou1}
    \fmfv{label=$p_{3/2}$,l.a=135}{b2}
    \fmfv{label=$\lambda 2^+$,l.a=-90}{b1}
\end{fmfgraph*}}\right]
\right\}
$$
$$
(-1.59)
$$
\end{widetext}
By summing up the different contributions to $T$ (bracket numbers)
one obtains the spectroscopic factor $|T_{1/2^-}|^2$=0.32.
The same technique has been used to obtain $|T_{1/2^+}|^2$=0.27 (cf. Table I). 
The calculation of the spectroscopic factors associated with the reaction $^{10}$Be(d,p)$^{11}$Be displayed 
in Table \ref{tab1} was carried out following the same scheme.

We conclude that the main nuclear structure properties of both $^{11}$Be and $^{12}$Be 
may be understood in terms of the self-energy 
and induced interaction processes associated with the polarization of the nuclear medium. 
The similarity of $NFT$ results with those of large shell model calculations reported in ref. \cite{Nav.00} for $^{12}$Be
and in ref. \cite{Ots.93} for $^{11}$Be,
indicate that a proper treatment of single-particle and of collective degrees of freedom and of 
their interweaving provide an essentially
complete description of the nuclear structure of these nuclei as was already found in the case of nuclei
lying along the stability valley.

\end{fmffile}

\bibliographystyle{apsrev}

\end{document}